\documentclass[a4paper,11pt]{article}
\pdfoutput=1 

\usepackage{jcappub} 

\usepackage{amsmath}
\usepackage{amsfonts,amssymb}
\usepackage{dsfont}
\usepackage{hyperref}
\usepackage{graphicx}
\usepackage{cleveref}
\usepackage[utf8]{inputenc}
\usepackage{xcolor}
\usepackage{bm}
\usepackage{braket}
\usepackage{amsthm}

\usepackage{todonotes,multirow,enumitem}

\hypersetup{
    colorlinks, 
    linktoc=all, 
    linkcolor=red,  
  citecolor=hyptxt,
  urlcolor=blue}
  \hypersetup{
  citebordercolor=,
  filebordercolor=green,
  linkbordercolor=blue
}
\definecolor{hyptxt}{rgb}{0.7, 0.4, 0.9}
\title{\boldmath Bianchi IX gravitational collapse of matter inhomogeneities}

\author[a]{Leonardo Giani,}
\author[b,c]{Oliver F. Piattella,}
\author[d,e]{Alexander Yu. Kamenshchik}

\affiliation[a]{The University of Queensland, School of Mathematics and Physics,\\ QLD 4072, Australia}
\affiliation[b]{Dipartimento di Scienza ed Alta Tecnologia, Universit\'a dell’Insubria, Via Valleggio 11, I-22100 Como, Italy}
\affiliation[c]{Department of Physics, Universidade Federal do Esp\'irito Santo, Avenida Fernando Ferrari 514, 29075-910 Vit\'oria, Esp\'irito Santo, Brazil}
\affiliation[d]{Dipartimento di Fisica e Astronomia, Università di Bologna and INFN, Via Irnerio
46,40126 Bologna, Italy}
\affiliation[e]{L.D. Landau Institute for Theoretical Physics of the Russian
Academy of Sciences, 119334 Moscow, Russia}

\emailAdd{uqlgiani@uq.edu.au}

\abstract{We investigate a model of gravitational collapse of matter inhomogeneities where the latter are modelled as Bianchi type IX (BIX) spacetimes. We found that this model contains, as limiting cases, both the standard spherical collapse model and the Zeldovich solution. We study how these models are affected by small anisotropies within the BIX potential. For the spherical collapse case, we found that the model is equivalent to a closed FLRW Universe filled with matter and two perfect fluids representing the anisotropic contributions. From the linear evolution up to the turnaround, the anisotropies effectively shift the value of the FLRW spatial curvature, because the fluids have effective Equation of State (EoS) parameters $w \approx -1/3$. Then we estimate the impact of such anisotropies on the number density of haloes using the Press-Schechter formalism.
If a fluid description of the anisotropies is still valid after virialization, the averaged over time EoS parameters are $w\approx 1/3$. Using this and demanding hydrostatic equilibrium, we find a relation between the mass $M$, the average radius $R$ and the pressure $p$ of the virialized final structure.
When we consider within the BIX ansatz small deviations from the Zeldovich solution, our qualitative analysis suggests that the so called \textit{pancakes} exhibit oscillatory behavior, as would be expected in the case of a vacuum BIX spacetime.}

\begin{document}
\maketitle
\flushbottom
\section{Introduction}
The cosmological principle, roughly defined\footnote{See for example Refs. \cite{Wald:1984rg,Weinberg:1972kfs} for a rigorous mathematical definition.} as the assumption that the Universe is homogeneous and isotropic on sufficiently large scales, is one of the pillars of the cosmological standard model $\Lambda$CDM.
On the other hand, anisotropies and inhomogeneities are of the utmost importance when we focus our attention on smaller scales. 
Indeed, whilst the average density of the Universe is close to the critical one (roughly  $10^{-26} kg/m^3$, i.e. few hydrogen atoms per cubic meter), most of the matter we observe is organized in denser clusters confined within the boundaries of large Dark Matter (DM) haloes, connected by filaments and surrounded by voids.
These structures originated in the early Universe from fluctuations of the
Cosmic Microwave Background (CMB) average temperature of order $10^{-5} K$ \cite{Planck:2018vyg}, who resulted in gravitational potential wells from which matter fall or escape.
Eventually, some of these regions become dense enough to trigger the gravitational collapse and the formation of the Large Scale Structures (LSS) we observe today. 

However, the description of the gravitational collapse of these matter inhomogeneities with analytical methods becomes difficult once we enter in the nonlinear regime. Some physical intuition is usually achieved starting from simplifying symmetry assumptions, which allow an analytical treatment.
One such an example is the Top Hat Spherical Collapse (THSC) model, which describes the collapse of an initially slight overdense spherical shell of nonrelativistic matter \cite{1967ApJ...147..859P,1972ApJ...176....1G,Mota:2004pa}. Already such a simple model accurately predicts some statistical properties of the haloes, like the number of expected collapsed objects of radius $R$ and mass $M$ which are formed in a time $t$ once a Gaussian distribution for the initial overdensities is assumed, see the seminal works by Bond and Meyers \cite{1996ApJS..103....1B} and Press and Schechter \cite{Press:1973iz}, or Ref. \cite{Weinberg:2008zzc} for a pedagogical introduction to the formalism.

On the other hand, the hypothesis of a perfectly spherical symmetric collapse may be unrealistic, and it is sensible to ask what happens when this assumption is relaxed. To this extent, one important extension of the Press-Schechter work based on a model of ellipsoidal gravitational collapse was proposed by Sheth and Tormen \cite{Sheth:1999su}, resulting in a better agreement with high resolution N-Body simulations. The ellipsoidal collapse model has been studied extensively in the literature, see for example Refs. \cite{Angrick:2010qg,Ludlow:2011jx,Kuhnel:2016exn,Suto:2016zqb}.
From the analytical point of view, an interesting description of anisotropic collapse is given by the Zeldovich solution, see Refs. \cite{Zeldovich:1969sb,1972A&A....20..189S,Pauls:1994pp}, which describes the gravitational collapse triggered by a 1-dimensional overdense perturbation of an otherwise flat FLRW spacetime.
The Zeldovich solution predicts the formation of  2-dimensional structures usually referred to as \textit{pancakes}, whose first observation by the Very Large Array (VLA) \footnote{\hyperref[https://www.cv.nrao.edu/nvss/]{https://www.cv.nrao.edu/nvss/}} was reported in Ref. \cite{1991PhRvL..67.3328U}.

On the other hand, most of the cosmic web is composed by filaments, i.e. 1-dimensional structures which must have generated from 2-dimensional anisotropic collapse. The latter cannot be described analytically neither within the ellipsoidal collapse or the Zeldovich solution, which motivate us to explore more general forms of gravitational collapse.
Since our goal is to describe an inhomogeneity which is initially expanding with the background, and then detaches from it and begin to collapse, it is reasonable to demand that the geometry of such inhomogeneity is spatially \textit{closed}. Indeed, for ordinary matter, an initially expanding overdensity with open or flat topology may cease to expand only asymptotically.\footnote{This was demonstrated in \cite{Wald:1983ky}, and generalized to inhomogeneous and anisotropic spacetimes in \cite{Kleban:2016sqm}.  } Furthermore, it is sensible to ask that its evolution can be parametrized in terms of the same time parameter we use to describe the flat FLRW background, i.e. it is possible to use the cosmic time to label the evolution of the inhomogeneity, which implies that the latter is \textit{homogeneous}.
There are eleven different homogeneous but anisotropic spacetimes, classified by E. Bianchi in Ref. \cite{L.Bianchi}. However, amongst them, the only one spatially closed is the Bianchi IX. The dynamics of the latter has been studied extensively in the past, and led to the discovery of the oscillatory approach to the cosmological singularity, see Refs. \cite{doi:10.1080/00018737000101171,doi:10.1080/00018738200101428,Belinski:2017fas}.
With this motivation in mind, we  explore the possibility of using the Bianchi IX geometry as a toy model to describe the 2-dimensional anisotropic collapse of a matter inhomogeneity. Since the Bianchi IX model contains as limiting cases, as we will show later, both the spherical collapse and the Zeldovich solution, it may describe within the same framework the evolution of filaments, pancakes and spherical objects composing the cosmic web. In this work we assess the impact of small anisotropies, constrained by the BIX potential, on the THSC model and the Zeldovich solutions.
We found that already for those simple cases interesting cosmological implications arise. In the former, for example, we show that the anisotropies can be modeled as if they were barotropic perfect fuids on a spherically symmetric background. This allowed us to study their general behavior during the expansion of the inhomogeneity up to the turnaround, which results in a modification of the Press-Schechter Halo mass function. We also speculate on the expected qualitative behavior the inhomogeneity should have during the collapse and at virialization. Regarding the Zeldovich solution, we found that the Bianchi IX potential triggers a dynamical force against an unbounded growth of anisotropy along a specific direction, transforming the 1-dimensional collapse in a 2-dimensional one. 

For applications concerning general inhomogeneities with arbitrary values of $\beta_\pm$, one should choose suitable junction conditions for the embedding of the Bianchi IX inhomogeneity within the isotropic Universe. This is certainly non-trivial, and a rigorous investigation of the topic deserves further studies which go beyond the scope of this preliminary work and we will address in a forthcoming paper.

In the past years there has been a growing interest in tests of the cosmological principle, which have been carried using a variety of sources.
Some analysis, based on  type Ia Supernovae, found no statistical evidence of violation of the cosmological principle, see for example Refs.\cite{Deng:2018jrp,Bengaly:2015dza,Andrade:2018eta,Wang:2017ezt,Javanmardi:2015sfa,Cai:2013lja}. On the other hand, other analysis point toward different results, see for example Refs. \cite{Zhao:2013yaa,Schwarz}. Other tests based on the distribution and luminosity of galaxies and galaxy clusters have a similar outcome, some works, see for example Refs.\cite{Bengaly:2015xkw,Appleby:2014lra}, found no statistically significant evidence against the cosmological principle, whilst other, see for example Refs.\cite{Colin:2017juj,Bengaly:2017slg,Migkas:2020fza,Migkas:2021zdo}, claim the opposite. Furthermore, investigations using different probes like the distribution of quasars and the distribution of gamma-ray burst indicate deviations from the Cosmological principle, see Refs.\cite{Secrest_2021,Balazs:1998tt}.
In particular, X-ray measurements of the scaling relations of galaxy clusters, \cite{Migkas:2020fza, Migkas:2021zdo}, detected an apparent $9\%$ spatial variation of the local rate of expansion $H_0$ across the sky. Interestingly, this is of the same order of magnitude of the statistical tension on the value inferred for it using early and late times cosmological probes, see for example Refs.\cite{Bernal:2016gxb, Verde:2019ivm} for an overview of the issue.
A mechanism through which the local Universe acquires an intrinsic anisotropy may be, therefore, an useful tool to investigate the aforementioned observations. Speculatively, we suggest that the final stage of the evolution of a Bianchi IX inhomogeneity, once we appeal to some virialization mechanism, may be a suitable candidate.  
We further speculate that, since the Bianchi IX spacetime symmetry is the non-Abelian rotation group $SU(2)$, it could also have interesting implications for spin properties of DM haloes and LSS.

The structure of the paper is the following: in Sec. \ref{model} we introduce the model and its generic features, in Sec. \ref{ASC} we study how small anisotropies modify the spherical collapse, whilst in Sec. \ref{ZSP} how they affect the Zeldovich solution. Finally, Sec. \ref{Conclusions} is devoted to a summary and a discussion of our results.

\section{The model} \label{model}
\subsection{Bianchi IX field equations}

It is known (see Ref. \cite{Landau:1982dva}) that in setting up the Einstein field Equations for Bianchi Universes there is no need of use explicit expressions for the basis vectors as function of coordinates.
From the following line element:
\begin{equation}
    ds^2 = -dt^2 + a^2(t) \omega^1\otimes\omega^1+b^2(t) \omega^2\otimes\omega^2+
c^2(t) \omega^3\otimes\omega^3 \;,
\end{equation}
where $a, b$ and $c$ are scale factors and $\omega^1,\omega^2$ and $\omega^3$ 
are the Maurer-Cartan basis 1-forms for the BIX spacetime:

\begin{eqnarray}
\omega_1 &=& -\sin{x_3}dx_1 + \sin{x_1}\cos{x_3}dx_2\;, \\
\omega_2 &=& \cos{x_3}dx_1 + \sin{x_1}\sin{x_3}dx_2\;,\\
\omega_3 &=& \cos{x_1}dx_2 +dx_3\; , 
\end{eqnarray}
see for example  Refs. \cite{RyanShepley+2015,Landau:1982dva}, we obtain the following non-vanishing components of the four dimensional Ricci tensor:
\begin{align}\label{00BIX}
R^0_0 = \left(\frac{\ddot{a}}{a} + \frac{\ddot{b}}{b}+\frac{\ddot{c}}{c}\right)\; ,
\end{align}
\begin{align}\label{11BIX}
R^1_{1}= \frac{\ddot{a}}{a} + \frac{\dot{a}}{a}\left(\frac{\dot{b}}{b} + \frac{\dot{c}}{c}\right) +\frac{\left(a^2 + b^2 - c^2 \right)\left(a^2 +c^2 - b^2\right)}{2a^2b^2c^2} \; ,\\\label{22BIX}
R^2_{2}= \frac{\ddot{b}}{b} + \frac{\dot{b}}{b}\left(\frac{\dot{a}}{a} + \frac{\dot{c}}{c}\right) +\frac{\left(b^2 + a^2 - c^2 \right)\left(b^2 +c^2 - a^2\right)}{2a^2b^2c^2} \; , \\\label{33BIX}
R^3_{3}= \frac{\ddot{c}}{c} + \frac{\dot{c}}{c}\left(\frac{\dot{b}}{b} + \frac{\dot{a}}{a}\right) +\frac{\left(c^2 + b^2 - a^2 \right)\left(c^2 +a^2 - b^2\right)}{2a^2b^2c^2} \; .
\end{align}

Let us consider a dust perfect fluid with Energy momentum tensor:
\begin{equation}
    T^{\mu}_{\nu} = \rho\left(u^\mu u_{\nu}\right)\;,
\end{equation}
where $u^{\mu}$ is the four velocity satisfying $u^{\mu}u_{\mu}=-1$. We will further assume that in our reference system the dust fluid is at rest, so that $u^i = 0$.
Then we can write down the Einstein field equations:
\begin{equation}\label{EFE}
    R_{\mu\nu} = T_{\mu\nu} - \frac{1}{2}g_{\mu\nu}T \;.
\end{equation}

It is convenient to express the scale factorswith the following parametrization by Misner \cite{Misner:1969hg}:
\begin{align}
    a(t) &= e^{\Omega + \frac{\beta_+}{2} + \frac{\sqrt{3}}{2}\beta_-} \; , \\
    b(t) &= e^{\Omega + \frac{\beta_+}{2} - \frac{\sqrt{3}}{2}\beta_-} \; , \\
    c(t) &= e^{\Omega - \beta_+} \; ,
\end{align}
where $\Omega$ is related to the volume (i.e. $abc = e^{3\Omega}$), and $\beta_{\pm}$ parametrize deviations from isotropy. Therefore, $e^{\Omega}$ is the ``average radius'', or the radius that the Universe would have without anisotropy. 

The field equations using these variables become:
\begin{align}\label{FriedmannequationBIX}
    \dot{\Omega}^2 &= \frac{\rho}{3} + \frac{1}{4}\left(\dot{\beta}_+^2 + \dot{\beta}_-^2\right) + \frac{\mathcal{K}}{3}e^{-2\Omega}\;,\\ \label{acceqBIX}
    \ddot{\Omega} &= -\frac{\rho}{2} - \frac{3}{4}\left(\dot{\beta}_+^2 + \dot{\beta}_-^2\right) - \frac{\mathcal{K}}{3}e^{-2\Omega} \; ,  \\
\label{beta+eq}
    \ddot{\beta}_- &+ 3\dot{\Omega}\dot{\beta}_- + \mathcal{K}_{\beta_-}(\beta_+,\beta_-,\Omega)=0  \;, \\
    \ddot{\beta_+} &+ 3\dot{\Omega}\dot{\beta}_++\mathcal{K}_{\beta_+}(\beta_+,\beta_-,\Omega)=0 \; ,\label{beta-eq}
\end{align}
where we have defined the effective spatial curvature\footnote{Note that in the limit $\beta_\pm \rightarrow 0$, $\mathcal{K}\rightarrow -3/4e^{-2\Omega}$, i.e. a standard closed FLRW curvature term with radius of curvature $r= 2a$, see also Ref. \cite{Landau:1982dva}} $\mathcal{K}$:
\begin{equation}
\mathcal{K} = \frac{1}{4}\left(e^{-4\beta_+} + e^{2\beta_+ - 2\sqrt{3}\beta_-} -2e^{-\beta_+ +\sqrt{3}\beta_-}-2e^{-\beta_+ -\sqrt{3}\beta_-}-2e^{2\beta_+} +e^{2\beta_+ +2\sqrt{3}\beta_-}\right)\; ,
\end{equation}
as well as the anisotropic curvature terms $\mathcal{K}_{\beta_\pm}$:
\begin{align}
    \mathcal{K}_{\beta_+}= \frac{1}{3}e^{-2\Omega}\left(-2e^{-4\beta_+} - 2e^{2\beta_+} +e^{2\beta_+ -2\sqrt{3}\beta_-} + e^{-\left(\beta_+ -\sqrt{3}\beta_-\right)}+ e^{-\left(\beta_+ +\sqrt{3}\beta_-\right)} +e^{2\beta_+ +2\sqrt{3}\beta_-}\right) \; ,
    \end{align}
    \begin{align}
    \mathcal{K}_{\beta_-}=\frac{e^{-2\Omega}}{\sqrt{3}}\left(e^{-\left(\beta_+ + \sqrt{3}\beta_-\right)} +e^{2\left(\beta_+ + \sqrt{3}\beta_-\right)}-e^{-\left(\beta_+ -\sqrt{3}\beta_-\right)}-e^{2\left(\beta_+ - \sqrt{3}\beta_-\right)}\right) \; ,
\end{align}
$\mathcal{K}_{\beta_\pm} = \frac{2}{3}e^{-2\Omega} d\mathcal{K}/d \beta_\pm$.
Taking the time derivative of Eq.\eqref{FriedmannequationBIX} and  noticing that:
\begin{equation}
    \mathcal{K}_{\beta_\pm} = \frac{2}{3}e^{-2\Omega} d\mathcal{K}/d \beta_\pm\;,
\end{equation}
we can combine \cref{acceqBIX,beta+eq,beta-eq} to write down the continuity equation for the dust fluid:
\begin{equation}
    \dot{\rho} + 3\dot{\Omega}\rho = 0\;,
\end{equation}
which shows that the density correctly dilute with the volume of the inhomogeneity.
It is possible to map the Bianchi IX field equations in those for a closed FLRW Universe filled with two, non-minimally coupled and interacting scalar fields, see Appendix \ref{appendixA} for the details. As we will show later, when $\beta_\pm \ll 1$ the two scalar fields decouple and results in two independent Klein-Gordon equations.

\subsection{Relation with other analyitical models of gravitational collapse}

It is interesting to note that Eqs. \eqref{FriedmannequationBIX}\eqref{acceqBIX} already contain, as limiting case, two well known analytical models of gravitational collapse, i.e. the THSC and the Zeldovich solution.

The THSC model describes the Newtonian evolution of a uniformly overdense spherical shell of an otherwise flat and dust-filled FLRW spacetime.
The Euler equation for such a shell of radius $a$, see for example chapter 6 of Ref.\cite{Mukhanov:2005sc}, becomes:
\begin{equation}
    \ddot{a} = -a\frac{4}{3}\pi G\rho_{sh}a \;,
\end{equation}
where $\rho_{sh}=\rho_{bg}\left(1+\delta\right)$ is density of the shell, where $\rho_bg= \rho_{0}a^{-3}$ is the density of the background matter field and $\delta$ the relative constant overdensity of the shell.
The latter equation has the following first integral:
\begin{equation}\label{THSCEq.}
    \frac{\dot{a}^2}{a^2} = \frac{8\pi G}{3}\rho_{sh} + \frac{K}{r^2} \;,
\end{equation}
where $K$ is an integration constant fixed by the initial conditions. Choosing these in such a way that initially the perturbation is small and the shell follows the background evolution one can relate $K$ with the initial strenght fluctuation, see chapter 8 of Ref.\cite{Weinberg:2008zzc} for the general method including a cosmological constant $\Lambda$.
The continuity equation of the shell, together with Eq. \eqref{THSCEq.} are nothing else than the Friedmann equations for a closed matter dominated Universe.
As we already mentioned, choosing $\beta_\pm = 0$ in Eqs. \eqref{FriedmannequationBIX} equivalent to those of a closed FLRW Universe, which therefore describe the THSC.

The Zeldovich solution gives the evolution of a 1-dimensional perturbation of a flat FLRW spacetime with line element:
\begin{equation}
    ds^2 = -dt^2 + a^2(t)\left(1-\lambda(t)\right)dx^2 + a^2(t)\left(dy^2 + dz^2\right) \;.
\end{equation}
The field equations for the latter become, see for example section (6.4) of Ref. \cite{Mukhanov:2005sc}:
\begin{equation}\label{zeldacceq}
    \dot{H} + H^2 = -\frac{\rho_{hom}}{6}\;,
\end{equation}
\begin{equation}\label{zeldlambda}
    \ddot{\lambda} +2H\dot{\lambda} - 4\pi G \rho_{\hom}\lambda = 0\; ,
\end{equation}
where $\rho = \rho_{hom}\left(\lambda - 1\right)^{-1}$ and $\rho_{\hom}= \rho_0 a^{-3}$.
Note that Eq. \eqref{zeldlambda} is, formally, identical to the one for the density contrast of non-relativistic matter in linear perturbation theory. 
It is easy to show that eqs. \eqref{FriedmannequationBIX}, \eqref{acceqBIX} reduce to \eqref{zeldacceq} and \eqref{zeldlambda} when we set $\beta_- = 0$ and define the new variables $\alpha = e^{\Omega + \beta_+/2}$, $H = \dot{\alpha}/\alpha= \dot{\Omega} + \dot{\beta_+}/2$ and $\lambda = 1- e^{-3\beta_+/2}$.


\section{Impact of small anisotropies on the spherical collapse} \label{ASC}
To cast the Bianchi IX equations in a form similar to the corresponding ones for a FLRW background, let us define:
\begin{equation}
    \Omega \equiv \log R\;, \qquad \dot\Omega = \frac{\dot R}{R} \equiv H\;.
\end{equation}
In this definition, $R$ is some sort of average scale factor.
If we assume the $\beta$'s to be small, we can write the following Taylor expansions for $\mathcal{K},\mathcal{K}_{\beta_\pm}$:
\begin{equation}
    \mathcal{K} = \frac{1}{4}\left[-3 + 6\beta_+^2 + 6\beta_-^2 + O(\beta^3)\right]\;,
\end{equation}
\begin{equation}
    \mathcal{K}_{\beta_\pm}=\frac{2}{R^2}\beta_\pm +O(\beta^2)\;.
\end{equation}
Notice that the linear terms of the Taylor expansion in $\mathcal{K}$ simplifies, so that only quadratic terms in $\beta_\pm$ appears in  Eqs. \eqref{FriedmannequationBIX} and \eqref{acceqBIX}. The same does not happen for the expansion of $\mathcal{K}_{\beta_\pm}$, where the linear term suffices, as expected from $\mathcal{K}_{\beta_\pm} \propto \partial \mathcal{K}/\partial\beta_\pm$. Interestingly, in terms of the multiscalar field description given in \ref{appendixA}, this implies that $V\left(\Omega, \varphi_\pm\right)$ decouples into two separable potential terms for the scalar fields and a standard geometrical spatial curvature contribution.

We then have:
\begin{align}
    H^2 &= \frac{\rho}{3} - \frac{1}{4R^2} + \frac{1}{4}\left(\dot{\beta}_+^2 + \dot{\beta}_-^2\right) + \frac{1}{2R^2}\left(\beta_+^2 + \beta_-^2\right)\;,\\ 
    \dot{H} &= -\frac{\rho}{2} + \frac{1}{4R^2} - \frac{3}{4}\left(\dot{\beta}_+^2 + \dot{\beta}_-^2\right) - \frac{1}{2R^2}\left(\beta_+^2 + \beta_-^2\right)\; ,  \\
    \ddot{\beta}_\pm &+ 3H\dot{\beta}_\pm + \frac{2}{R^2}\beta_\pm = 0\;. 
\end{align}
In this way we can treat the problem as a usual spherical collapse for dust where the anisotropy,  parametrized by the $\beta$'s, is described as an effective fictitious homogeneous and isotropic fluid. In this framework we can assess, at least on average, how anisotropy thwarts or enhances the collapse. Note that, usually, when a component with pressure is added to dust in order to study their conjoint collapse, the condition of ``top hat'', i.e. a ``step'' profile for the energy density, cannot be maintained because pressure gradients do not allow this. In our present case, however, the extra component is a fictitious one; so, even if it possesses pressure, as we are going to see, this should not invalidate the hypothesis of a ``top hat'' profile be maintained during the collapse. 

Let us define:
\begin{equation}
    \rho_{\beta_\pm}= \frac{3}{4}\dot{\beta}_\pm^2 + \frac{3}{2}\frac{\beta_\pm^2}{R^2}\;,
\end{equation}
then the first Friedmann equation becomes:
\begin{equation}
    H^2 = \frac{\rho + \rho_{\beta_+} + \rho_{\beta_-}}{3} - \frac{1}{4R^2} \;.
\end{equation}
The extra, fictitious component is evident here. The acceleration equation becomes:
\begin{equation}
    \dot{H}^2 = -\frac{\rho}{6} + \frac{1}{4R^2} -\sum_{i=\pm}\left(\frac{3}{4}\dot{\beta}_i^2 +\frac{1}{2}\frac{\beta_i^2}{R^2}\right)\;,
\end{equation}
hence:
\begin{equation}
    p_{\beta\pm}=\frac{3}{4}\dot{\beta}_\pm^2 - \frac{\beta_\pm^2}{2R^2}
\end{equation}
It can also be checked that, differentiating $\rho_{\beta_\pm}$ and using 
\begin{equation}
    \ddot{\beta}_\pm + 3H\dot{\beta}_\pm + \frac{2}{R^2}\beta_\pm = 0\;,
\end{equation}
one obtains the correct continuity equations:
\begin{equation}
    \dot\rho_{\beta_\pm} = -3H(\rho_{\beta_\pm} + p_{\beta_\pm})\;.
\end{equation}
The equation of state parameter for the $\beta_\pm$ components are then:
\begin{equation}\label{betaeos}
    w_{\beta_\pm} = \frac{\frac{3}{4}\dot{\beta}_\pm^2 - \frac{\beta^2_\pm}{2R^2}}{\frac{3}{4}\dot{\beta}_\pm^2 + \frac{3}{2}\frac{\beta^2_\pm}{R^2}}\;,
\end{equation}
so, it is even a ``respectable'' component, in the sense that it is never phantom $w_\beta < -1$ or super-stiff $w_\beta > 1$.

\subsection{Linear growth}
Demanding that initially the volume of the almost spherical BIX spacetime follows the background matter dominated FLRW evolution, we expect $H \approx 2/3t$. Inserting this in the Klein-Gordon Equation for $\beta_\pm$ we obtain:
\begin{equation}
    \ddot{\beta}_\pm + \frac{2}{t}\dot{\beta}_\pm + \frac{2}{a_0^2}t^{-\frac{4}{3}}\beta_\pm = 0\;.
\end{equation}
The general solution of the latter equation is a combination of Bessel functions of order $1/2$, which we can write as:
\begin{equation}
    \bar{\beta}(t) = \frac{\bar{\beta}_0}{t}\cos\left(\omega t^{\frac{1}{3}} + \psi_0\right) +\frac{\bar{\beta}_1}{t^{\frac{2}{3}}}\cos\left(\omega t^{\frac{1}{3}} + \psi_1 \right) \;, 
\end{equation}
which, as expected, shows that small anisotropies generated during the linear evolution of the inhomogeneity oscillate and are smoothed out by the cosmological expansion. Neglecting the $\bar{\beta}_0$ mode, which decays faster, we are left with $\bar{\beta} (t)\approx \bar{\beta}_1\cos{\left(\omega t^{1/3} + \psi\right)}a^{-1}$. 
Since in this regime the kinetic energy of the scalar field decays as $\dot\beta^2 \propto t^{-3}$, while the potential energy goes as $\beta^2/R^2 \propto a^{-4} \propto t^{-8/3}$, we can conclude that eventually the scalar fields become potential dominated.
As a result, in this approximation, the Eos parameter of the anisotropic fluid  from Eq. \eqref{betaeos} is $w_{\beta_\pm}\approx -1/3$.
Therefore, the continuity equations for the $\rho_{\beta_\pm}$ fluids become:
\begin{equation}
    \frac{\dot{\rho}_{\beta_\pm}}{\rho_{\beta_\pm}}=-3H\left(1+w\right)= -2H\;,
\end{equation}
whose solutions are:
\begin{equation}\label{rhobetalinear}
    \rho_{\beta_\pm} \propto R^{-2}\;.
\end{equation}
The latter result shows that, during the linear stage of the evolution, the effects of the anisotropic fluids on the averaged spherical collapse is to effectively shift the value of the spatial curvature term. This is not unexpected, as studied in detail in Refs. \cite{Kamenshchik:2011jq,Khalatnikov:2019uyx}, and results from the rich phenomenology offered by a perfect fluid with Eos $w= -1/3$ in Friedmann and spherically symmetric spacetimes. 

\subsection{Turnaround}
Even if initially the inhomogeneity follows the background evolution, it will eventually slow down  and cease its expansion.
The turnaround point is reached when the volume of the inhomogeneity $e^{3\Omega}$ reaches its maximum value, and therefore $H=0$. 
The Klein-Gordon equations in this regime become:
\begin{equation}
    \ddot{\beta}_\pm + \frac{2}{R^2}\beta_\pm = 0 \;, 
\end{equation}
i.e. the equation for a standard harmonic oscillator whose solution is:
\begin{equation}
    \beta_\pm = \beta_{0\pm}\cos\left(\omega t + \psi\right) \;,
\end{equation}
where $\omega = \sqrt{2}/R$. 
Using the latter solution in the Friedmann Equation we obtain:
\begin{equation}\label{turnaround}
    R^2 = \frac{3}{\rho}\left(\frac{1}{4} - \frac{1}{2}\beta_{0+}^2 -\frac{1}{2}\beta_{0-}^2\right) \;.
\end{equation}

The latter result, together with the solution \eqref{rhobetalinear} we obtained for the linear evolution, confirms that the anisotropic fluids at these stages of the evolution of the inhomogeneity effectively reduce the absolute value of the spatial curvature term.
This ultimately slows down the detachment from the Hubble flow compared to the spherical case with same initial conditions for the scalar perturbations strength. As a result, the turnaround will generally happens at higher average radius. This may seems in contradiction with Eq. \eqref{turnaround}, but in the latter we have to keep in mind that if the expansion phase is longer, $\rho$ will generally dilute more. To see that explicitly, let us look at the analytical solutions of the Friedmann equations for the spherical collapse. These can be expressed in a parametric form, when the spatial curvature is negative, as:
\begin{eqnarray}
R(\theta)_{sph} = \frac{4\pi G \rho_0}{3|K|}\left(1-\cos\theta \right)\;,\qquad
t(\theta)_{sph} = \frac{4\pi G\rho_0}{3|K|^{\frac{3}{2}}}\left(\theta - \sin\theta\right)\;,
\end{eqnarray}
where $|K|$ is the value of the spatial curvature. For our model, when $\beta_\pm =0$, $|K| = 1/4$. On the other hand, the presence of small anisotropies effectively shifts the value of $|K|$ to $|\tilde{K}|= 1/4 - \rho_{\beta_+^0} - \rho_{\beta_-^0}= |K|\left(1-4\rho_{\beta_+}^0 - 4\rho_{\beta_-}^0\right)$. As a consequence we have, at first order:
\begin{eqnarray}\label{t-rBIX}
R(\theta)_{BIX} = R(\theta)_{sph}\left(1+4\rho_{\beta_+}^0+4\rho_{\beta_-}^0\right)\;, \qquad
t(\theta)_{BIX}= t(\theta)_{sph}\left(1+6\rho_{\beta_+}^0 + 6\rho_{\beta_-}^0\right)\;,
\end{eqnarray}
which clearly shows that the turnaround happens later, and at higher radius than in the spherical case.

\subsection{Contraction and virialization}
When the inhomogeneity evolves from the turnaround point to the contracting phase, i.e. when $H<0$, the Klein-Gordon equations for the $\beta$'s possess an anti-damping term, which will eventually lead to the instability of the anisotropies. 
On the other hand, even in the standard top hat spherical collapse the density $\rho$ of the inhomogeneity shall at some point become unstable,  unless we appeal to a somehow \textit{ad hoc} chosen virialization mechanism.
Furthermore, the geometry of the vacuum Bianchi IX spacetime prevents the anisotropy to grow indefinitely because of the triangular potential wells, against which the system would eventually bounce off.

If virialization happens soon enough for the $\beta$'s to remain small and in such a way that the two uncoupled fluids description still holds, we can assess the ``averaged'' impact of the anisotropy on the virialized structure. 
Indeed, let us assume that the anisotropic fluids are in hydrostatic equilibrium with the gravitational field of the virialized halo. Then their pressure and densities are related as:
\begin{equation}\label{Newhydreq}
    \frac{d p_{\beta_\pm}}{dR} = -g(R)\rho_{\beta_\pm} \;,
\end{equation}
where $g(R)$ is the strenght of the gravitational field. 
Since the $\beta$'s are small, let us suppose that $g(r) \approx M/R^2$, where $M = \int d V \rho(R)$, i.e let us neglect the anisotropic field's densities contribution to the total mass. 
Since after virialization the volume of the halo becomes constant, we have $H=0$. The Klein-Gordon equations solutions are therefore simply harmonic oscillators, and we can evaluate the anistropic fluid Eos as:
\begin{equation}\label{Eosbetasmall}
    w_{\beta_\pm}= 1 - \frac{4}{3}\frac{\beta_{\pm}^2}{\beta_{0\pm}^2} = 1 - \frac{4}{3}\cos^2\left(\omega t + \psi\right) \;.
\end{equation}
From  Eq.\eqref{Eosbetasmall} we see that the equation of state parameter for the anisotropic fluid is oscillating and bounded $-1/3 \leq w \leq 1$. On the other hand, if we use the averaged value of $<cos^2> = 1/2$ in Eq.\eqref{Eosbetasmall}, we get $w= 1/3$, i.e. the anisotropic fluids behave as radiation. Using the latter in Eq. \eqref{Newhydreq} we obtain:
\begin{equation}
    \frac{d p_{\beta_\pm}}{p_{\beta_\pm}} = 3\frac{M}{R^2}dR \;.
\end{equation}
Assuming that the mass through the Halo is uniformly distributed (i.e. does not depends on $R$), the latter equation is then straightforwardly integrated and gives:
\begin{equation}\label{betapressurevir}
    p_{\beta_\pm}(R_{vir}) = e^{3\frac{M}{R_{vir}}}.
\end{equation}
Eq.\eqref{betapressurevir} allow us to quantify, once that the mean radius of a virialized halo $R_{vir}$ as well as his mass $M_{tot}$ are known, its averaged (over time) pressure due to the presence of the anisotropic fluids.

We must stress however that the result of Eq. \eqref{betapressurevir} strongly depends on the assumption that the anistoropic fields $\beta_\pm$ could still be described as perfect fluids. This may be unrealistic since we know that virialization is induced by the interactions between the particles which compose the fluid, with the result of converting the internal energy of the latter into orbital, stationary motions preventing the collapse. On the other hand, the anistropic fluids $\beta_\pm$ are fictitious and therefore do not contain real particles, so that no interaction could take place between them. As a result, we believe that our picture describes how the stationary orbits between the matter particles are deformed on average by the anisotropies.

\subsection{Impact on statistical large scale structures observables}

To understand how our model modifies the statistical distribution of LSS we will make use of the Press-Schechter formalism \cite{Press:1973iz}. For a scale invariant power spectrum of primordial scalar fluctuations ($n_s = 1$), this model predicts that the number density $n$ of haloes with mass between $M$ and $M+\Delta M$ is given by:
\begin{equation}
n\left(M,t\right)dM = \frac{\bar{\rho}}{M^2}\sqrt{\frac{2}{\pi}}\left|\frac{d\log{\nu}}{d\log M}\right|\nu e^{\frac{-\nu^2}{2}}\;,
\end{equation}
where $\bar{\rho}$ is the background matter density and it was defined $\nu=\delta_c/\sigma_M$, in which  $\sigma_M$ is the mass variance and $\delta_c = 1.686 D(t)$, with $D(t)$ the linear growth factor normalized to unity.
The value of $\delta_c/D(t) = 1.686$ is a prediction of the standard top hat spherical collapse, and is given by the ratio $(\rho(t_c) - \bar{\rho}(t_c))/\bar{\rho}(t_c)$, i.e. the surplus of matter within the inhomogeneity (compared to the background) that would be present at the collapse time $t_c$ if the linear theory would still hold.
The collapse time $t_c$, since the solution is periodic, is simply  $t_c=2t_{ta}$.
Remembering that the density contrast during matter domination is:
\begin{equation}
    \delta = \frac{3}{20}\left(\frac{6\pi t}{t_{ta}}\right)^{\frac{2}{3}}\;,
\end{equation}
and using  Eqs. \eqref{t-rBIX} we have, at first order:
\begin{equation}
    \delta = \frac{3}{20}\left(\frac{6\pi t}{t_{ta_{sph}}}\right)^{\frac{2}{3}}\left(1-4\rho_{\beta_+}^0 -4\rho_{\beta_-}^0 \right) \;.
\end{equation}
The latter equation shows that when $t=2t_{ta_{sph}}$, we have $\delta = 1.686\left(1-4\rho_{\beta_+}^0 -4\rho_{\beta_-}^0\right)$.
Thus, within these effective approximations, the effect of small anisotropies of the spherical collapse is to rescale the function $\nu$ of a factor $\nu \rightarrow\tilde{\nu}=\nu\left(1-4\rho_{\beta_+}^0 -4\rho_{\beta_-}^0\right)= \kappa\nu$. Accordingly, the number density becomes:
\begin{equation}\label{numberdensitybix}
 n\left(M,t\right)dM = \frac{\bar{\rho}}{M^2}\sqrt{\frac{2}{\pi}}\left|\frac{d\log{\nu}}{d\log M}\right|\kappa\nu e^{-\frac{\left(\kappa\nu\right)^2}{2}}\;.
\end{equation}

  \section{Impact of small anisotropies on the Zeldovich solution}\label{ZSP}
 
 The Zeldovich equations \eqref{zeldacceq},\eqref{zeldlambda} under the assumption $\beta_- \ll 1$ become:
 \begin{equation}
     \dot{H} + H^2 = -\frac{\rho_{hom}}{6} \; ,
 \end{equation}
  \begin{equation}\label{2orderlambdaeq}
      \ddot{\lambda} + 2 H\dot{\lambda} -\frac{\rho_{hom}}{2}\lambda -\frac{3\dot\beta_{-}^2}{2}\left(1-\lambda\right)=0\;.
  \end{equation}
 The Klein-Gordon equation for $\beta_-$ becomes, at linear order:
\begin{equation}\label{bmkgeqzel}
    \ddot{\beta}_- + \dot{\beta}_-\left(3H - \frac{\dot{\lambda}}{1-\lambda}\right) + \frac{2\beta_-}{\alpha^2}\left( \frac{2}{\left(1-\lambda\right)^2}-1\right)=0\;.
\end{equation}
We do know that, neglecting the term $\dot{\beta}^2_-$ in Eq. \eqref{2orderlambdaeq}, the solution for $\lambda$ in the matter dominated epoch is the same as the density contrast for nonrelativistic matter, i.e. $\lambda = \lambda_0t^{2/3}\propto \alpha$.

Using the latter to evaluate the coefficient of the velocity term in Eq. \eqref{bmkgeqzel} we obtain:
\begin{equation}
    \left(3H - \frac{\dot{\lambda}}{1-\lambda}\right) = H\left(\frac{3-4\lambda}{1-\lambda}\right)\;.
\end{equation}
Let us also suppose that the background is strongly anisotropic, i.e. $\lambda \gg 1$. In this regime, Eq. \eqref{bmkgeqzel} simplifies as:
\begin{equation}
    \ddot{\beta}_- + \frac{8}{3t}\dot{\beta}_- 
    -\frac{2 }{\alpha_0^2 t^{\frac{4}{3}} }\beta_-=0 \;,
\end{equation}
which admits the following real solution:
\begin{equation}
\beta_- =
    \frac{10}{18}\bar{\beta}_{-}\frac{ \left(\left(\frac{1}{3 k t^\frac{2}{3}}+1\right) \sinh \left(3 \sqrt{k} t^{\frac{1}{3}} \right)-\frac{ \cosh \left(3 \sqrt{k} t^{\frac{1}{3}}\right)}{\sqrt{k} t^{\frac{1}{3}}}\right)}{k^\frac{3}{2}t} \;,
\end{equation}
where $k= 2/\alpha_0^2$ and $\bar{\beta}_-$ is an integration constant. The latter solution is unstable and implies that the growing $\beta_-$ will at some point spoil the validity of the perturbative approach. To get a qualitative understanding of how the evolution of $\lambda$ is affected, let us rewrite Eq. \eqref{2orderlambdaeq} for $\lambda \gg 1 $:
\begin{equation}
    \ddot{\lambda} +2H\dot{\lambda} \approx \lambda\left(\frac{\rho_{hom}}{2}-\frac{3}{2}\dot{\beta}_-^2\right) \;.
\end{equation}
It is straightforward to realize that, as soon as $3\dot{\beta}_-^2$ becomes bigger than $\rho_{hom}$, an effective force appears working against the growth of $\lambda$. 
We can conclude that the growth of  $\beta_-$ triggers the appearance of a dynamical force against the original perturbation $\lambda$.
This qualitative picture is not surprising, and it is in agreement with what would we expect for a vacuum Bianchi IX spacetime. Indeed in the latter, for general initial conditions, anisotropy along one direction cannot grow arbitrarily  because of the triangular shape of the potential $V(\beta_+,\beta_-)$, so that the system will eventually bounce from one of the potential wells and change the direction of anisotropic contraction.

This result suggests that, within the Bianchi IX model of gravitational collapse for structure formation, the so called \textit{pancakes} of the Zeldovich solution are deformed by the switching of the direction of contraction and expansion, and undergo oscillatory behavior.
     
\section{Summary and discussion}\label{Conclusions}
One of the goals of Cosmology is to explain how the LSS we observe today evolved from the highly, but not perfectly, homogeneous and isotropic primordial Universe. This, in turns, requires a good understanding of the physics ruling the evolution of the initially small perturbations of the matter power spectrum. During the linear evolution of the inhomogeneities, where things are still analytical and relatively simple, our understanding is excellent. Unfortunately, the highly nonlinear nature of the EFE unavoidably introduces complications when the perturbative approach fails. To develop intuition, it seems logical to start with reasonable simplifying assumptions about the geometry of these inhomogeneities, and exploit their symmetries to make the EFE easier to handle.  In our opinion, one of such reasonable but quite general assumption is that, no pun intended, these inhomogeneities are, by themselves, homogeneous. It is sensible to demand, indeed, that their individual evolution can be parametrized by mean of their own ``time'' parameter.
The THSC, for example, is based on the assumption that the inhomogeneity is in fact a closed FLRW spacetime.
With this motivation in mind, in this work we studied the evolution of the primordial inhomogeneities under the assumption that they have the geometry of the BIX. The latter emerges as the most natural candidate within the Bianchi classification of homogeneous spacetimes, since it is the only one which is topologically closed and recovers in the limit of vanishing anisotropy the closed FLRW spacetime. 

 Let us briefly recapitulate the most interesting results presented here before discussing their implications:
\begin{itemize}
    \item The BIX geometry contains as limiting cases both the THSC model (in the trivial case of vanishing anisotropy) and the Zeldovich solution for a 1-dimensional perturbation. This provides a common framework to describe spherical DM haloes and Zeldovich pancakes.    
    \item For almost spherical inhomogeneities, before the collapse,  our qualitative analysis shows that the anisotropies effectively change the value of the FLRW spatial curvature. The reason is that the anisotropic fields in this regime, curiously, mimic a fluid with an EoS parameter $w\approx-1/3$ (see Eq. \eqref{rhobetalinear}), whose energy density is degenerate with the  one of the spatial curvature.
    \item Still assuming small deviations from sphericity, we studied how the anisotropies affect the number density of collapsed objects, see Eq.\eqref{numberdensitybix}. They result in a rescaling of the Gaussian peak of the distribution and of the total number of objects, while the mass dependence is unchanged.
    \item If at virialization the anisotropies are still small and can be described as perfect fluids, we found that their EoS parameters oscillate between $w= -1/3$ and $w=1$. However, their mean value (averaged over time) is $w\approx 1/3$. Demanding hydrostatic equilibrium between these fluids and the gravitational field of the virialized object, we obtain the pressure profile of Eq. \eqref{betapressurevir} at the boundaries of the almost spherical final structure.
    \item To get a flavor of the behaviour of the model beyond the assumption of almost sphericity, we studied anisotropic deviations from the Zeldovich solution. These turn out to be unstable, and work against the growth of the original 1-dimensional perturbation. This does not come as a surprise, since the full unperturbed BIX spatial curvature potential prevents anisotropy from growing unbounded in a specific direction.\footnote{With the exception, of course, of the singular point, where however the very notion of anisotropy becomes debatable.} We do know, however, that the very same potential should trigger a new, qualitatively very similar, differently oriented 1-dimensional growth (which in vacuum would correspond to a Kasner epoch).
    Therefore, our qualitative understanding is that the final stage of the evolution are not the pancakes, but more complex objects evolved from a series of subsequent Zeldovich-like epochs before virialization. These, to us, seem a promising tool to mimic the rich variety of filaments weaving the cosmic web.    
\end{itemize}

As we discussed, the Bianchi IX spacetime is a natural candidate for modelling the inhomogeneities of the primordial matter power spectrum, once we require that they are closed and homogeneous. In this work, we devote most of our attention to the behaviour of the model around the closed FLWR and Zeldovich solutions, which are unstable critical points of the BIX spacetime. The analysis shows that, even in these simple cases, a rich phenomenology arises,and highlights the potential of our proposal.  Most of our conclusions about the evolution beyond the approximation of small perturbations are quite speculative because of the complexity of the BIX potential. Nevertheless, we believe that these speculations are reasonable because of the qualitative behavior of BIX when the spatial curvature becomes dominant. 
Finally, a very tickling speculative question we would like to ask is the following: what if the final structure, after virialization, has inherited the internal symmetries of the BIX spacetime? Since the latter is the non Abelian group $SU(2)$, the resulting DM Haloes could possess intrinsic spin and open a window towards new, charming phenomenology. 

For all these reasons, we believe that the BIX geometry is a promising tool towards a better understanding of the physics of structure formation, and deserves further investigations.

\acknowledgments
We are grateful to Tamara M. Davis for valuable comments and discussions. LG acknowledges support from the Australian Government through the Australian Research Council Laureate Fellowship grant FL180100168, AK is partially supported by the Russian Foundation for Basic Research grant No. 20-02-00411,  OFP acknowledges CNPq (Brazil) for partial financial support.

\appendix
\section{Description in terms of a multiscalar tensor theory}\label{appendixA}

One of the advantages of working within the  $\Omega$ and $\beta_\pm$ coordinates is that it is possible to map the vacuum Bianchi IX spacetime Einstein Field Equations (EFE) in those for a flat FLRW Universe filled with two interacting and non minimally coupled scalar fields (which may be helpful for numerical analysis).
To see this equivalence, let us rewrite explicitly the Einstein Hilbert action for the Bianchi IX spatial geometry within the ADM formalism \cite{Arnowitt:1962hi}:

\begin{equation}
S = \int N(t) dt d^3x \sqrt{g}\left(K_{ij}K^{ij} - K^2 + R_3\right)\;,   
\end{equation}
where $g_{ij}$ is the spatial metric with the 3-dimensional Ricci scalar $R_3$, $N(t)$ is the lapse function and $K_{ij}=(\partial_t g_{ij}/2N)$.
In terms of the variables $\Omega, \beta_\pm$  the latter action reads:
\begin{equation}\label{ADMaction}
    S = \int dtd^3x \sqrt{-g}\left[\frac{1}{N}\left(-6\dot{\Omega}^2+ \frac{3}{2}\dot{\beta}_+^2 + \frac{3}{2}\dot{\beta}_-^2 \right) + 2Ne^{-2\Omega}\mathcal{K}\right]\;.
\end{equation}
Varying the above action with respect to $N$ gives the Friedmann Equation \eqref{FriedmannequationBIX}. We can then choose the gauge $N=1$ to recover the synchronous reference system, and vary with respect to $\Omega, \beta_\pm$ to obtain the remaining field equations.

In the action \eqref{ADMaction}, since  $\sqrt{g}_{BIX}=\sqrt{g}_{FLRW}=e^{3\Omega}$,  we can easily identify the scalar Lagrangian for a flat FLRW spacetime defining the scale factor $e^{\Omega}=a$ :
\begin{equation}
    R_{FLRW}=  6\left(\frac{\ddot{a}}{a} +\frac{\dot{a}^2}{a^2}\right) = \left(6\ddot{\Omega} + 12\dot{\Omega}^2\right) \simeq -6\dot{\Omega}^2\;,
\end{equation}
where the symbol $\simeq$ means equivalent up to total derivative term.
Therefore, defining the new variables $\varphi_\pm = \sqrt{3}\beta_\pm$, we can write down the full action as:
\begin{equation}
    \int d^4x\sqrt{-g} \left(R_{FLRW} + \mathcal{L}_{\varphi_{\pm}}\right) \;,
\end{equation}
where the scalar fields Lagrangian is given by:
\begin{equation}
    \mathcal{L}_{\varphi_\pm} = \frac{\dot{\varphi}_\pm^2}{2} + V_{int}\left(\varphi_\pm,\Omega\right)\;,
\end{equation}
and the interaction potential reads: 
\begin{equation}
    V_{int}\left(\varphi_\pm,\Omega\right)= 2e^{-2\Omega}\mathcal{K}\;.
\end{equation}

Varying with respect to the scalar fields we obtain the  Klein-Gordon equations:
\begin{equation}
    \ddot{\varphi}_\pm + 3\dot{\Omega}\dot{\varphi}_\pm + V_{,\varphi_\pm} = 0 \;,
\end{equation}
with $V_{,\varphi_\pm}$ denoting the functional derivative of $V$ with respect to $\varphi_\pm$.

\bibliographystyle{apsrev4-2.bst}
\bibliography{TDD.bib}
\end{document}